\documentclass{aa}
\usepackage{graphicx} 

\begin{document} 
 
\thesaurus{06(02.01.2; 02.09.1; 08.02.1; 08.14.2)} 
 
\title{Z~Cam stars: a particular response to a general phenomenon} 
\titlerunning{Z~Cam: a particular response of a general phenomenon} 
\authorrunning{Buat-M\'enard et al.} 
\author{Valentin Buat-M\'enard\inst{1}, Jean-Marie Hameury\inst{1} 
        \and Jean-Pierre Lasota\inst{2}} 
\offprints{V. Buat-M\'enard} 
\mail{buat@astro.u-strasbg.fr} 
 
\institute{UMR 7550 du CNRS, Observatoire de Strasbourg, 11 rue de 
           l'Universit\'e, F-67000 Strasbourg, France, 
           hameury@astro.u-strasbg.fr \and Institut d'Astrophysique de 
           Paris, 98bis Boulevard Arago, 75014 Paris, France, 
           lasota@iap.fr} 
\date{Received / Accepted} 
 
\maketitle 
 
\begin{abstract} 
 
We show that the disc instability model can reproduce all the observed 
properties of Z~Cam stars if the energy equation includes heating of the 
outer disc by the mass-transfer stream impact and by tidal torques and if 
the mass-transfer rate from the secondary varies by about $30\%$ around 
the value critical for stability. In particular the magnitude difference 
between outburst maxima and standstills corresponds to observations, all 
outbursts are of the inside-out type and can be divided into two 
classes: long (wide) and short (narrow) outbursts, as observed. Mass 
transfer rate fluctuations should occur in other dwarf novae but one can 
exclude variations similar to those observed in magnetic systems (AM 
Her's and some DQ Her's) and some nova-like systems (VY Scl's), in which 
$\dot{M}$ become very small during low states; these would produce 
mini-outburst which, although detectable, have never been observed. 
 
\keywords accretion, accretion discs -- instabilities -- (Stars:) 
novae, cataclysmic variables -- (stars:) binaries : close 
\end{abstract} 
 
\section{Introduction} 
 
Dwarf novae are a subclass of cataclysmic variable stars, which are 
close binary systems in which matter transferred from a Roche-lobe 
filling secondary star is accreted by a primary white dwarf (see 
\cite{warner95}). In dwarf novae accretion proceeds 
through a disc which is the site of more or less regular outbursts. 
The recurrence time of these 2-6 magnitude outbursts can range from 10 
days to several tens of years. The dwarf nova outbursts are attributed 
to a thermal-viscous instability occurring in the accretion disc. 
Although the thermal-viscous disc instability model (DIM) identifies a 
physical cause for the instability, partial hydrogen ionization, the 
outburst cycles it produces differ in many important respects from the 
observed ones (see e.g. Smak 2000, Lasota 2000). It is now rather 
widely accepted that the original DIM has to be completed by various 
physical processes such as irradiation of the disc and the secondary 
star, inner disc truncation and mass transfer fluctuations. Also 
heating of the disc by the mass-transfer stream impact and by tidal 
torque dissipation plays an important role in the dwarf-nova outburst 
cycle (Buat M\'enard et al., hereafter Paper I). 
 
Dwarf novae can be divided into three subclasses: U Gem-type stars 
which have the most regular outburst cycles; SU UMa-type stars showing 
both short and very long outbursts (superoutbursts) and Z~Cam stars. 
The last group is characterized by a `standstill' phenomenon: the 
decline from outburst maximum is interrupted and the luminosity of the 
systems settles to a value $\sim 0.7$ mag lower than the peak 
luminosity. In some cases the magnitude difference is smaller (see e.g. 
Fig. 19 in Lin, Papaloizou \& Faulkner 1985). Such standstills may 
last from ten days to years. After that, the system luminosity declines 
to the usual quiescent state. In his pioneering work, Osaki (1974) 
interpreted standstills as stable phases of accretion in the framework 
of the disc instability model he then proposed. In his disc-radius and 
mass-transfer rate diagram, Smak (1983) described these stars as an 
intermediate case between stable nova-like stars and unstable dwarf 
novae. Meyer \& Meyer-Hofmeister (1983) proposed that Z~Cam stars are 
dwarf novae with a mass-transfer rate that fluctuates about the 
critical rate. Lin et al. (1985) concluded that moderate fluctuation 
of the mass-transfer rate can produce Z~Cam-type light curves. This is 
now widely accepted, but until now no model has been able to reproduce 
the observed light curves of \object{Z~Cam} (e.g. \cite{king98}). 
 
We applied the disc instability model to a system in which the mass 
transfer rate from the secondary star $\dot{M}_2$ is assumed to vary 
around the critical rate $\dot{M}_{\rm c}$, above which the disc is 
stable. In Section 2 we show that the magnitude difference between 
standstills and outburst peaks depends on the difference between 
$\dot{M}_{\rm c}$ and the actual mass transfer rate reached during 
standstill $\dot{M}_{\rm h}$. We also show that it is easier to find a 
large enough $\Delta_{\rm mag}$ when the model includes the outer disc 
heating by the stream impact and the tidal dissipation as discussed in 
Paper I. In addition, including these additional heating processes in 
the DIM allows to obtain inside-out outbursts as observed in some Z~Cam 
stars.  We show therefore that in such a generalized DIM the response of a 
dwarf nova disc to low-amplitude mass-transfer rate variations is the 
explanation of the Z~Cam phenomenon. 
 
Mass-transfer rate variations are also occurring in other dwarf novae 
(e.g. Smak 1999, 2000; Hameury, Lasota \& Warner 2000). The brightening 
of the hot spot, the region where the mass-transfer stream interacts 
with the outer disc's rim, which is observed near the outburst maxima, 
should be the result of a mass-transfer rate enhancement. Large 
variations of the mass transfer rate are directly observed in polars, 
which are magnetic cataclysmic variables with no accretion discs. The 
prototype, \object{AM Her} shows large amplitude variations of the mass 
transfer rate which can be vanishingly small during low states. Similar 
luminosity drops observed in some DQ Her stars, magnetic non-synchronous 
binaries and in VY Scl, nova-like systems with orbital periods between 3 
and 4 hours, are also attributed to mass-transfer fluctuations. However, 
because of the presence of a disc, it is much more difficult to convert 
these luminosity fluctuations into mass-transfer rate variations. Contrary to 
Schreiber et al. (2000), who used a code in which the outer boundary 
radius is kept fixed, we find using the correct outer boundary 
condition, that such drastic mass-transfer fluctuations do not occur in 
most dwarf novae. King \& Cannizzo (1998) reached the same conclusion as 
ours using a fixed outer radius and varying the mass-transfer rate by 6 orders 
of magnitude. 
 
\section{Modeling the Z~Cam phenomenon} 
 
\subsection{Model and parameters} 
 
We use the version of the disc instability model described in Hameury et 
al. (1998). We assume that $\alpha = \alpha_{\rm cold} = 0.04$ in 
quiescence and $\alpha = \alpha_{\rm hot} = 0.2$ in outburst. As in 
Paper I the fixed, inner radius is at $r_{\rm in} = 10^9$ cm. We allow for
the inclusion of the stream impact and tidal dissipation effects 
(see Paper I for details). We do not include the effects of the 
secondary star irradiation, i.e. the mass-transfer rate variations are 
supposed, in this model, to be unrelated to the accretion onto the white 
dwarf. 
 
We choose \object{Z~Cam} as the system to model. From the Ritter and Kolb's 
catalogue (1998) we took mass of the primary $M_1 = 1 M_\odot$, the mass of 
the secondary $M_2 = 0.7 M_\odot$ and the orbital period $P_{\rm orb} = 6.96$ 
hr. We use the tables from Lubow \& Shu (1975) and Paczy\'nski (1977) to 
estimate the circularization radius $r_{\rm circ} = 1.52 \times 10^{10}$ cm 
and the mean outer disc radius $<r_{\rm out}> = 5.4 \times 10^{10}$ cm. The 
size of the disc is controlled by the tidal torques which depend on the value 
of the circularization radius (see Hameury et al. 1998 and Paper I for 
details). 
 
We assume that the mass-transfer rate from the secondary varies from a 
low state $\dot{M}_{\rm l} = <\dot{M}> - \Delta \dot{M}$ to a high 
state $\dot{M}_{\rm h} = <\dot{M}> + \Delta \dot{M}$ with a period 
$P_{\rm m}$, where $<\dot{M}>$ is the average mass-transfer rate from 
the secondary. Transitions between the two states are made by 
sinusoidal interpolation during a time interval $\Delta t_{\rm 
trans}$. In this paper we use $P_{\rm m} = 400$ days; this is long 
enough to account for relaxation during standstills and to allow 
several outbursts cycles. The transition time is taken to be $\Delta 
t_{\rm trans} = 20 $ days. 
 
The brightening of the quiescent level and the outburst amplitude 
reduction observed in some Z~Cam systems (Szkody \& Mattei 1984) may 
result from a particular time-dependence of the mass-transfer rate 
fluctuations but we are not attempting to model such effects here. 
The $\dot{M}$ variations used in this paper are not supposed to 
correspond to any particular physical mechanism. We are interested in 
the disc's response to mass-transfer rate fluctuations and not in the 
possible causes of these variations. The latter are far from being 
known. Irradiation induced cycles of mass-transfer rate variation are 
excluded in this context as they last much longer than the observed 
interval between standstills (\cite{gon93}). Star spots could be be 
responsible for $\dot{M}$ fluctuations that are relevant for Z~Cam 
stars (\cite{livio94}; \cite{king98}). 
 
\subsection{Mass-transfer fluctuations} 
 
\begin{figure} 
\resizebox{\hsize}{!}{\includegraphics{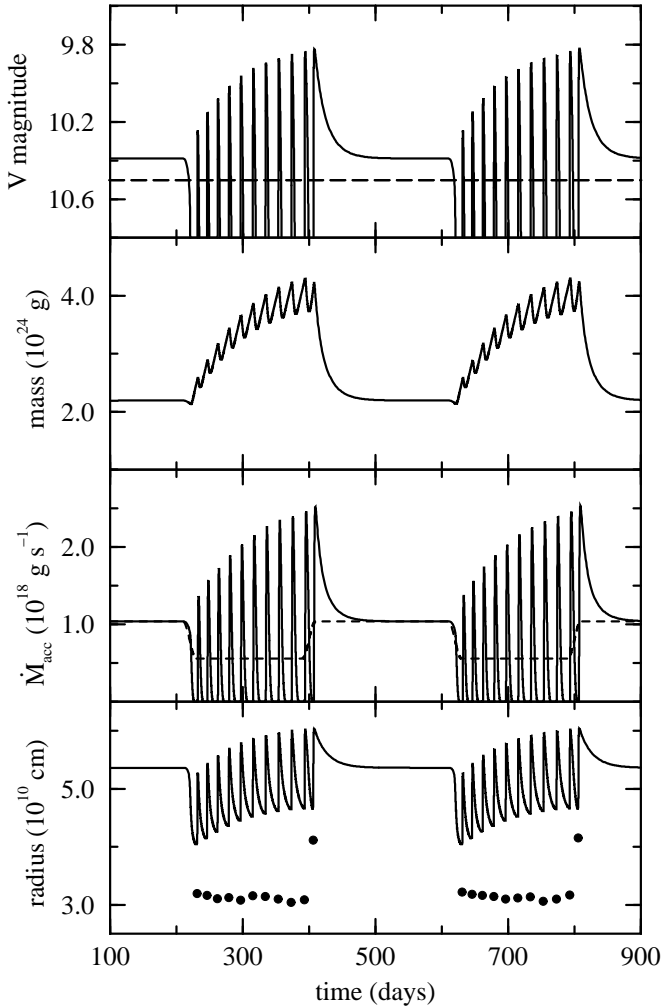}} 
\caption{Results of the standard DIM for the parameters of \object{Z 
Cam}, assuming a modulated mass-transfer rate $8.0 \times 10^{17} \pm 30 
\%$ g s$^{-1}$. The upper panel shows the visual magnitude (solid line). 
The dashed line represents the V magnitude (10.5) of a disc which 
accretes at exactly the critical value: $\dot{M}_{\rm c} = 8.0 \times 
10^{17}$g s$^{-1}$. The second panel from above shows variations of the 
disc's mass. The third one represents the mass accretion rate onto the 
white dwarf (solid line), as well as the mass transfer rate from the 
secondary (dashed line). The bottom panel shows the variations of the 
outer disc radius. The dots show point where the instability is 
triggered at the onset of an outburst. All outbursts are of the 
outside-in type.} 
\label{fig1} 
\end{figure} 
 
Fig. \ref{fig1} shows the results of our calculations when no additional
disc heating is included. We assumed the average mass-transfer rates
$<\dot{M}>$ to be equal to the critical rate $\dot{M}_{\rm c} = 8 \times
10^{17}$ g s$^{-1}$ (see Hameury et al. 1998 for the formula). We took
$\Delta \dot{M}/<\dot{M}> = 30\%$. The modulation of the mass-transfer
rate occurs at $t=200 + k P_{\rm m}$ day. As can be seen, the outburst
amplitudes gradually increase after a standstill, as the disk relaxes
towards the new situation in which the mass transfer rate is reduced.
Since now mass accumulates in the disc, it is more massive than before,
when it was steady. It is also larger after an initial contraction phase
due to the passage of a cooling front. All outbursts are of the
outside-in type, as the mass transfer rate is large (see Paper I). The
instability sets in at approximately the same position ($\sim 3 \times
10^{10}$ cm, except for the last outburst which is triggered by an
increase of the mass transfer rate from the secondary). As only a small
fraction of the disk mass is accreted during outbursts, a relaxed regime
is reached only after a large number of outbursts, close to the end of
the dwarf nova phase. The luminosity increase of successive outbursts, a
characteristic feature of this model, is not observed in reality. To
solve this problem one could use severals remedies. For example, one
could add to the model mass-transfer rate enhancements due to the
irradiation of the secondary by the accretion flow (Smak 1999; Hameury
et al. 2000). We will show below, however, that this is not necessary
since taking into account the heating by the stream impact and/or tidal
torque dissipation (Paper I) gives very satisfactory results. In any
case Fig. \ref{fig1} shows that $\Delta_{\rm mag} < 0.6 $ mag, which is
too low since the observed magnitude difference often exceeds 0.7 mag.
To get a better result would require the use of some fine tuning. For
obvious reasons, the standstill luminosity cannot be lower than that
corresponding to $\dot{M}_{\rm c}$, i.e. for our parameters the
brightest standstill magnitude can only be $m_V=10.5$. The outburst peak
luminosity depends mainly on the disc radius, which only weakly depends
on the mass-transfer rate. Therefore the magnitude at maximum, in our
case $m_V=9.8$ is just an invariant of the model. Therefore to obtain
the observed $\Delta_{\rm mag} \sim 0.7 $ would require $\dot{M}_{\rm h}
\simeq \dot{M}_{\rm c}$. Fine tuning would be required for the $\sim$30
Z~Cam known, because the critical and the peak accretion rate have the
same $r$-dependence. 
 
Another problem with our parameters is that the required vicinity to 
the critical rate implies very high mass-transfer rates. This produces 
outside-in outbursts in contradiction to observation. Moreover, according to 
the most recent work on CV's evolution \object{Z~Cam} should have a 
mass-transfer rate lower than $\sim 3.2 \times 10^{17}$ g s$^{-1}$ 
(Baraffe \& Kolb 2000). 
 
It is therefore possible to produce a Z~Cam-type light curve by varying
the mass-transfer rate around the value critical for disc stability, but
neither the parameters of this light-curve nor the required parameters
of the systems are satisfactory. One should also note that the range of
possible parameters is very restricted. While the mass-transfer rate is
$8 \times 10^{17}$ g s$^{-1}$ the accretion rate at maximum is only less
than a factor 2 to 3 larger ($1.3 - 2.5 \times 10^{18}$ g s$^{-1}$).
This means that a mass-transfer rate enhancement by a factor more than
$\sim$2 would produce a ``Z~Cam'' light-curve with $\Delta_{\rm mag} <
0$ as in King \& Cannizzo (1998), who increased the mass-transfer rate
by a factor 6. The reason for this very narrow range of mass-transfer
fluctuations which allow a Z~Cam-type standstill is that close to the
disc's outer edge non-local effects during the outburst bring the
surface density very close to the critical value $\Sigma_{\rm min}$ (see
e.g. Smak 1998, Fig. 4), i.e. $\dot M$ close to $\dot{M}_{\rm c}$. Since
at maximum the accretion rate in the disc is roughly constant
$\dot{M}_{\rm max} \sim \dot{M}_{\rm c}$. Esin, Lasota \& Hynes (2000)
assumed that $\dot{M}_{\rm max} = \dot{M}_{\rm c}$ and deduced that an
enhancement of mass transfer rate above $\dot{M}_{\rm c}$ will always
produce a standstill brighter than the outburst maximum. This is not
true because in reality the peak accretion rate is larger than
$\dot{M}_{\rm c}$ by a factor of about 2. Therefore King \& Cannizzo
(1998) guessed correctly that they had increased the mass-transfer rate
by an amount too large to get a Z Cam-type standstill. One should note,
however, that some low-mass X-ray binary transient systems show plateau
brighter than a preceding local maximum which could be due mass-transfer
enhancements by factors larger than 2 (see e.g. Esin et al. 2000). 

\subsection{Effects of heating by stream impact and tidal torque dissipation} 
 
We found in the preceding section that in order to produce credible Z
Cam light-curves one should lower the critical mass-transfer rate. One
needs therefore a physical mechanism which would produce such an effect
without lowering the peak luminosity. Heating of the disc by the impact
of the mass-transfer stream from the secondary and/or by tidal torque
dissipation (Paper I) could do the job: it has a stabilizing effect,
therefore allowing the disc to be dimmer during standstills, but does
not affect the outbursts properties near the dwarf-nova maximum because
(i) as mentioned earlier they depend only weakly upon the average mass
transfer rate, and (ii) the additional heating is small compared to the
energy released during outbursts. Moreover, heating of the outer disc
reduces both the value of the critical mass-transfer rate $\dot{M}_{\rm
c}$ and the value $\dot{M}_{\rm AB}$ for which the outburst type
changes; this makes outside-in outbursts possible for reasonably low
values of the mass transfer rate (paper I). The relative difference
$(\dot{M}_{\rm c} - \dot{M}_{\rm AB})/\dot{M}_{\rm c}$ is also slightly
reduced. It is thus possible to have also inside-out outbursts
during the unstable phase of Z Cam systems even for moderate
fluctuations of the mass-transfer rate. This could explain why
type B outbursts are observed in some Z Cam systems, while other would
rather have type A ones (Warner 1995).

In a recent paper, Stehle \& King (2000) argued that heating of the disc
by the stream impact is responsible for the standstill luminosity being
less than the outburst peak luminosity, provided that the amplitude
of the mass transfer fluctuations is large enough : they require that
the mass transfer is low enough during the unstable phase for heating by
the stream impact to be negligible, and large during standstill.
However, as clearly showed by our model calculations such constraints are
not necessary. In addition, Stehle \& King model would require very fine
tuning: the mass transfer fluctuations would have to be large enough for
heating by the stream impact to be, depending on the mass-transfer
fluctuation phase, either negligible or quite important, and yet the
mass transfer rate during standstills could not be larger than
approximately (1.5 -- 2) $ \dot{M}_{\rm c}$, in order to prevent the
standstill phase from being too bright. Obviously one would expect to
find systems in which both conditions are not satisfied and such systems
would have excessively bright standstills.
 
\begin{figure} 
\resizebox{\hsize}{!}{\includegraphics{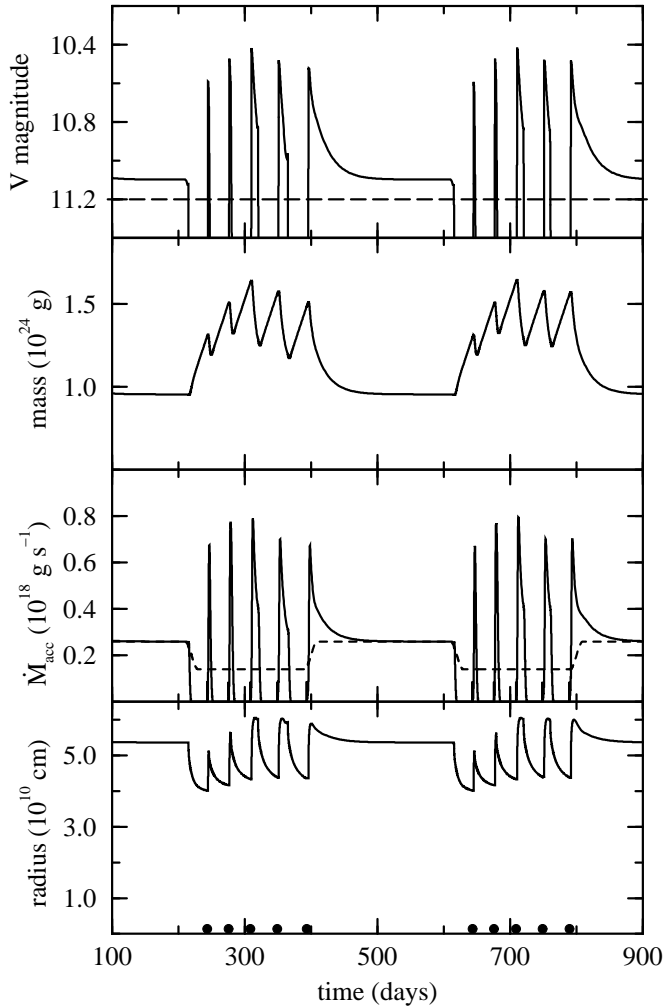}} 
\caption{Light curve obtained for Z~Cam parameters with the DIM 
including stream impact and tidal torques effects. $\dot{M}_2 = 2.0 
\times 10^{17} \pm 30 \%$g s$^{-1}$. The dashed line shows the 
standstill magnitude (11.2) for the critical mass transfer rate 
$\dot{M}_{\rm c} = 2.4 \times 10^{17}$g s$^{-1}$} 
\label{fig2} 
\end{figure}

Including only the heating of the disc by the stream impact (see
\cite{buat2000}) gives $0.55 < \Delta_{\rm mag} < 1.00$ mag for a fluctuation
of 30 \% around $<\dot{M}> = 3.0 \times 10^{17}$g s$^{-1}$.  The values of
both the mass transfer rate and the magnitude difference are thus in a better
agreement with the observed properties of Z~Cam stars.  Adding to this effect
the tidal dissipation term in the energy balance equation, gives $0.6 <
\Delta_{\rm mag} < 0.8$ mag for a fluctuation of 30 \% around $<\dot{M}> =
2.0 \times 10^{17}$g s$^{-1}$ (Fig.  \ref{fig2}), which is very satisfactory.
The cause of this success is a faster relaxation of the disc after the drop
in the mass-transfer rate, when both effects are included. This is due to two
effects. First, a larger mass fraction is accreted during an outburst:
$\Delta M / M \sim 0.25$. Second, the difference between the disc mass during
standstill and during the outburst phase is now smaller: additional heating
reduces the difference $\Sigma_{\rm max} - \Sigma_{\rm min}$ in the outer
regions of the disc (Paper I), where most of its mass sits. In addition, for
mass-transfer rates $<\dot{M}> \sim 2.0 \times 10^{17}$g s$^{-1}$ the two
external heating mechanisms produce narrow and wide outbursts (see
\cite{buat2000} for details). Such outbursts are observed but cannot be
reproduced by the standard DIM (Smak 2000). Our longer outbursts correspond
to the `plateau' outbursts of Oppenheimer et al. (1998).

In this particular model all outbursts inside-out type. In Z Cam and AH Her,
outbursts are presumed to be of type B whereas in RX And they are presumed to
be of type A (see table 3.6 in Warner (1995)). As mentioned above,  with the
inclusion of heating by stream impact and tidal torques, $(\dot{M}_{\rm c} -
\dot{M}_{\rm AB}) / \dot{M}_{\rm c}$ is reduced so that it is possible to
obtain type B outbursts even with a moderate mass transfer rate fluctuation. 
Depending on the characteristics of these fluctuations, one could obtain
either type A outbursts, or type B ones, or a mixture of both. In addition,
the last outburst of the unstable phase should be of type A if $\dot{M}
> \dot{M}_{\rm AB}$ before it starts. If the mass transfer rate varies
smoothly, one would therefore expect to observe type A outbursts just
before a standstill; this would not be the case if $\dot{M}$ varied
rapidly (for example if the secondary irradiation during outbursts
affects $\dot{M}$). 

As $\dot{M}_{\rm c}$ is close to $\dot{M}_{\rm AB}$ ( $\dot{M}_{\rm c} \sim
1.2 \dot{M}_{\rm AB}$ for Z Cam and SS Cyg), one should expect U Gem systems
in which outside-in outbursts occur, to have standstils from time to time.
Warner (1995) noted that the suggestion that ``all U Gem stars are
unrecognized Z Cam stars" is somewhat exagerated, but a number of U Gem stars
with short recurrence times have been reclassified as Z Cam stars. It would
therefore not be surprising if SS Cyg were to exhibit a standstill, even
though none has been observed in the past 100 years. This only requires a
slight increase of the mean mass transfer rate.
 
\begin{figure}
\resizebox{\hsize}{!}{\includegraphics[angle=-90]{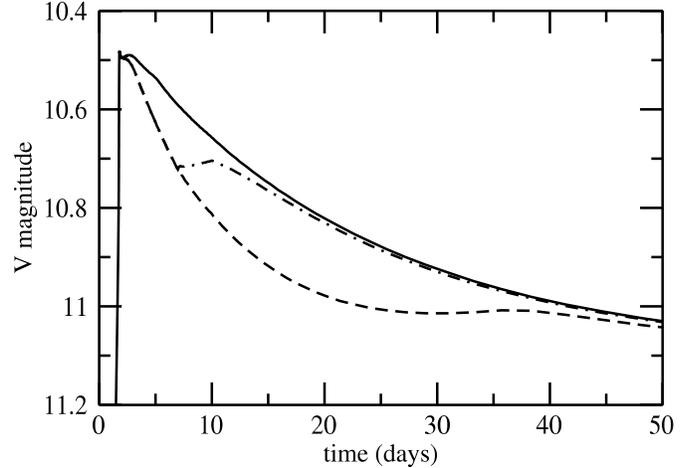}}
\caption{Last outburst before standstill. The solid line is an enlargement of
Fig. \ref{fig2}, for which mass transfer increases early in the outburst; the
dot-dashed curve is obtained when mass transfer starts increasing linearly 5
days after the outburst peak, and reached the standstill value 3 days later;
the dashed curved corresponds to a case where the increase in $\dot{M}$ also
begins 5 days after the outburst maximum, but with a timescale of 30 days.
The latter curve requires some tuning.}
\label{fignew}
\end{figure}

In general, despite the simplicity of its assumptions, our model represents
very well the outburst properties of Z~Cam stars. The recurrence time is 32
days between narrow outbursts and 40 days between plateau outbursts, the
duration of the two narrow outburst 8 and 9 days, of the two `plateau'
outbursts 16 and 20 days. These results compare reasonably well with the
observed values: the average cycle length of common and plateau outbursts is
23 and 31 days respectively, and their average duration 10 and 17 days. 
However, the duration of the last outburst before standstill is too long 
compared to observations:
we predict that it should be significantly longer than the previous ones,
whereas this is not the case for the observed outbursts. As a matter of fact,
the transition between the outburst phase and standstill depends significantly
on how $\dot{M}$ increases before standstill. 
Fig. \ref{fignew} shows different transitions to standstill, for the
same parameters as in Fig. \ref{fig2}, except that mass transfer now increases
during the outburst. As can be seen, shorter outbursts are easily obtained.
This might be used as an indication that $\dot{M}$ increases a few days after
the beginning of an outburst, maybe as the result of illumination; one must
however keep in mind that this is a relatively minor detail compared
to many uncertainties of the physics of the model (viscosity
prescription, etc.).

Z~Cam-type light curves can therefore be very well reproduced by the 
disc response to a simple mass-transfer rate fluctuations if additional 
heating is included.

\section{$\dot{M}_2$ variation in other dwarf novae} 
 
\begin{figure} 
\resizebox{\hsize}{!}{\includegraphics{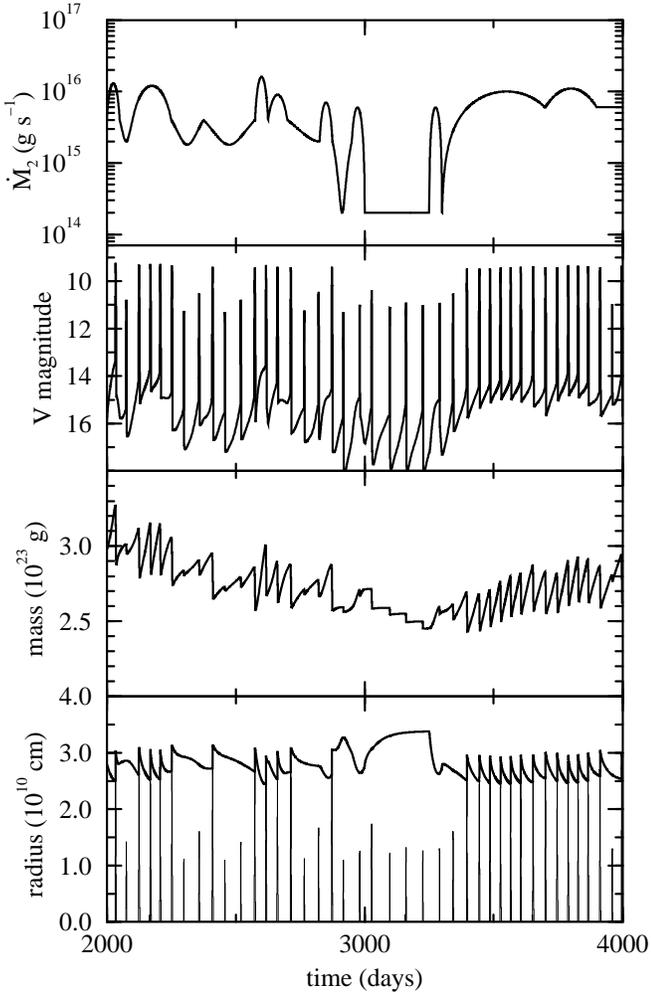}} 
\caption{DIM results for a fictitious dwarf nova with the parameters 
and the mass-transfer rate variation (top panel) of AM Her. The top 
panel gives the mass transfer rate from the secondary, the second 
panel the disc visual magnitude, the third panel its mass, and the 
bottom panel the outer disc radius (thick solid line), and the 
position of the external heating or cooling front (thin solid line).} 
\label{fig3} 
\end{figure} 
 
Variations of the mass transfer rate from the secondary are not
restricted to the Z~Cam subclass of dwarf novae. The variety of outburst
types observed in most dwarf-nova light curves implies the existence of
a free parameter for any given system, and the mass transfer rate from
the secondary is a good candidate; as discussed in Paper I the main
different outburst types (outside-in, inside-out, narrow and wide
outbursts) are triggered for different mass transfer rates. Finally, the
large luminosity variations of AM Her systems, CVs with no accretion
discs, must directly result from fluctuations of the mass transfer rate
from the secondary. Also in other systems, the magnetic non-synchronous
DQ Her stars and nova-like VY Scl stars, the large brightness variations
observed are attributed to mass-transfer fluctuations (see Garnavich \&
Szkody 1988 and references therein). 
 
However, the reasons for these fluctuations are not really known;
irradiation of the donor star and star spots have been invoked as
possible causes for the variations, but no reliable model has been yet
proposed. Moreover, the amplitude of these fluctuations remains, in the
general case, very uncertain. Smak (1995) studied the luminosity of the
hot spot in \object{Z Cha} and \object{U Gem} and showed that during
outburst the mass transfer rate can increase by a factor of about 2 ;
there was however no indication of very significant long term
variations. We have shown that for Z~Cam system a mass-transfer rate
variation of less than 50~\% (i.e. a factor less than 2) is required to
explain their light curves. On the other hand, Schreiber et al. (2000)
suggested that mass transfer rate variations similar to those observed
in \object{AM Her} could be present in dwarf novae systems. In such a
case, the mass transfer rate $\dot{M}$ could vary by one order of
magnitude and become vanishingly small during low states. This behaviour
was observed in synchronous magnetic systems, and could be quite
general; one must however point out again that there has been no report
of disappearance of the hot spot in dwarf novae, even if this might be
due to insufficient statistics. Two dwarf novae were observed in a low
state during quiescence: \object{HT Cas} and \object{BZ UMa}. BZ UMa was
found to be at 17.8 mag instead of the usual $\sim 16$ mag (Ka{\l}u\.zny
1986). BZ UMa, however, could be a DQ Her star (Kato 1999). In the 5
year of RoboScope observations, during which HT Cas was in quiescence,
long term (days to hundreds of days) $\sim 1.8$ mag variations were
observed. The quiescent luminosity varied from about 15.9 to 17.7 mag
(Robertson \& Honeycutt (1996)). During one of the low states Wood et al.
(1995) observed eclipsed X-ray emission which shows that accretion onto
the white dwarf was still going on. Although \object{HT Cas} is an
eclipsing system it is notorious for not showing the presence of a hot
spot, so there is no direct evidence that the low states are due to a
mass-transfer fluctuation of a corresponding size. The dwarf nova
\object{WW Ceti} shows 1 mag variations during quiescence, which could
be due to mass-transfer fluctuations (Ringwald et al. 1996). RoboScope
observations of eight well sampled SU UMa stars show no evidence for
large quiescent variations. The evidence for large mass-transfer
fluctuations in dwarf novae is therefore rather sparse. In any case, if
the observed variations are due to mass-transfer variations, they are
smaller than nova-like (or AM Her) variations: 1 to 1.8 mag against 2 to
6 mag. This is consistent with the conclusion that in dwarf novae such
huge variations can be excluded by the DIM, as we will show below. 
 
Schreiber et al. (2000) claim that although the cessation of mass 
transfer has immediate consequences on the light curves, these are 
limited to shortening the outburst duration and increasing 
the recurrence time. This is apparently in contradiction with the 
findings of King \& Cannizzo (1998) that the cessation of mass 
transfer results in a 2 mag decrease of the outburst amplitude; it 
must however be noted that the mass-transfer rate during low state was 
reduced by 6 orders of magnitude in King \& Cannizzo's models, but 
only by one order of magnitude in Schreiber et al. A major drawback 
of both sets of calculations, however, is the use of a fixed outer 
disc radius as a boundary condition; this severely affects the light 
curves, and is responsible for the occurrence of mini-outbursts at a 
constant $\dot{M}$ (\cite{jmh98}). 
 
To see what are the real predictions of the DIM we applied therefore 
our disc instability model to a fictitious dwarf nova in which the 
mass-transfer rate variations are similar to those of \object{AM Her} (see 
Fig. \ref{fig3}). We include stream impact and tidal dissipation for 
a dwarf nova with $M_1 = 0.6 $M$_\odot$, $M_2 = 0.26 $M$_\odot$, 
$P_{\rm orb} = 3.08$ hr and $r_{\rm out} = 2.7 \times 10^{10}$ 
cm. The mass-transfer rate varies from 0.01 up to $4 \times 10^{16}$g 
s$^{-1}$; the variations are similar to those in Schreiber et 
al. (2000). 
 
Three different types of modulations can be distinguished in the mass 
transfer curve: during the initial phase, $\dot{M}$ varies by a factor 
$\sim$ 10; then in some phases the mass transfer almost completely 
ceases, being reduced by a factor 100, during either a short or a long 
time; during the last 700 days, only small variations are present. When 
the mass transfer rate remains larger than a few $10^{15}$ g s$^{-1}$, 
the calculated light curve shows little variations; the outburst peak 
magnitudes do not change by more than 0.2 mag and of the recurrence time 
remains the same, to within a few day. When the mass transfer rate 
increases, so does the outburst peak magnitude with a delay of about ten 
days. By contrast, during the low state only mini-outbursts, fainter by 
2 mag than the normal ones, are present. The recurrence time increases 
by up to a factor 2. 
 
This is quite similar to the findings of King \& Cannizzo (1998) -- but 
our mass transfer rate is 4 orders of magnitude larger than theirs, and 
contradicts Schreiber et al. who consider the same variations of 
$\dot{M}$ as we do. The reason for such a discrepancy is the difference 
in the boundary conditions. During an outburst, the disc expands, so 
that tidal torques can carry the large amount of angular momentum of 
matter accreted onto the white dwarf. Under the effect of both the tidal 
torques and the mass transfered  from the secondary, the disc contracts during 
quiescence. When the mass transfer ceases, this contraction does no 
longer occur, and the disc remains extended, and the surface density is 
low. The outbursts are still of the inside-out type, but the heating 
front can no longer propagate into the outer regions where the disc 
surface density has been reduced below $\Sigma_{\rm min}$ under the 
effect of the previous disc expansion (see Fig. \ref{fig3}). The 
outburst amplitude is then very reduced. 
 
As noted by King \& Cannizzo (1998), the light curves of dwarf novae 
do not show such drastic variations of the outburst peak luminosity. 
Contrary to Schreiber et al. (2000), our results exclude drastic 
variations of mass transfer rate such as those observed in \object{AM Her}. 

\section{Conclusion} 
 
Since the work of Meyer \& Meyer-Hofmeister (1983) it has been believed
that the Z~Cam behaviour is due to mass transfer rate enhancements that
lead to phases of the outburst cycle during which the accretion disc is
steady (standstills). However, actual models produced light-curve whose
properties did not really correspond to observations (Lin et al. 1985;
\cite{king98}). In particular, it was not easy to reproduce the correct
magnitude difference ($\Delta_{\rm mag} \simeq$ 0.7 mag) between the
standstills and the outburst peaks and impossible to obtain inside-out
outburst with only moderate mass-transfer variations (Smak 1996). We
show that one can account for the observed $\Delta_{\rm mag}$ if the
amplitude of the mass transfer variations are not too large (typically
factors 2 or less), as suggested by Lin et al. (1985). An even better
agreement is obtained when one takes into account the energy released by
the mass-transfer stream impact onto the disc and by tidal torque
dissipation. Then the required mass transfer rates are also in much
better agreement with the expectations for these systems. The sequence
of long and short outbursts, the possibility to obtain inside-out outbursts 
and the characteristic time-scales of the light curves are also in very good
agreement with observations. 
 
There are no reasons to believe that mass-transfer rate variations would be 
restricted to some particular subclasses of dwarf-nova systems. On the 
contrary, the large variety of outburst shapes (duration and type, outside-in 
and inside-out) observed in the light curves can be explained by these 
fluctuations (\cite{jmh00}; \cite{buat2000}). Schreiber et al. (2000) argued 
that mass-transfer rate variations similar to those observed in \object{AM 
Her} should also be present in dwarf novae. However, if one uses correct 
outer boundary conditions, one obtains mini-outbursts fainter by about 2 
magnitudes during what would correspond to low states in AM Her systems. This 
has not been observed in dwarf novae; moreover, significant drops of the mass 
transfer rate should result in corresponding drops of the hot spot 
luminosity, which have not been observed either. In fact, in the few cases 
where mass-transfer rate variations in dwarf-nova systems seem to be observed 
these variations are order of magnitudes lower.

Similarly, if the mass transfer rate of nova-like systems were to undergo
variations similar to those observed in AM Her systems, most if not
all nova-like stars should exhibit a dwarf-nova behaviour, which obviously is not
the case. It is interesting to note that we know about 30 Z~Cam systems, among
several hundred CVs; the catalogue of Ritter and Kolb (1998) lists 16 Z~Cam
stars with known orbital periods, among a total of approximately 140 systems
in the same orbital period range, meaning that in the Z~Cam period range,
there is one chance out of 9 for a system to cross the stability limit,
indicating that the mass-transfer rate varies by a few tens of percent. 
 
The origin of these variations is not yet understood. It is known (Smak
1995, 1996) that during dwarf-nova outbursts, there is a significant
enhancement of the mass transfer rate; but this could not account for
erratic variations observed sometimes in quiescence. Stellar spots are a
good candidate for explaining these. Whatever the mechanism, it should
be influenced by the nature of the primary and/or the evolutionary
status of the secondary since we have shown that the pronounced low
states of the type observed in magnetic systems and in VY Scl cannot
exist in most of dwarf novae. 
 
\begin{acknowledgements} 
The authors wish to thank the referee, Prof. Y. Osaki, for his very
interesting comments and inspiring questions about the outburst types of
Z Cam systems. This work was supported in part by a grant from {\sl
Programme National de Physique Stellaire} of the CNRS. 
\end{acknowledgements}

\end{document}